\documentstyle[preprint,aps]{revtex}
\begin{document}
\title{In-plane polarized collective modes in detwinned YBa$_{2}$Cu$_{3}$O$_{6.95}$
observed by spectral ellipsometry}
\author{C. Bernhard$^{1}$, T. Holden$^{1}$, J. Huml{\'\i}{\v c}ek$^{2}$, D. Munzar$%
^{2}$, A. Golnik$^{3}$, M. Kl\"{a}ser$^{4}$, Th. Wolf$^{4}$, L. Carr$^{5}$,
C. Homes$^{5}$, B. Keimer$^{1}$, and M. Cardona$^{1}$}
\address{1) Max-Planck-Institut f\"{u}r Festk\"{o}rperforschung, Heisenbergstrasse 1,%
\\
D-70569 Stuttgart, Germany.\\
2) Institute of Condensed Matter Physics, Masaryk\\
University, Kotl\'{a}\v{r}sk\'{a} 2, CZ-61137 Brno, Czech Republic.\\
3) Institute of Experimental Physics, Warsaw University, Ho\.{z}a 69, 00-681%
\\
Warsaw, Poland.\\
4) Forschungszentrum Karlsruhe, ITP, D-76021 Karlsruhe, Germany.\\
5) National Synchrotron Light Source, Brookhaven National Laboratory, USA.}
\date{\today}
\maketitle

\begin{abstract}
The in-plane dielectric response of detwinned YBa$_{2}$Cu$_{3}$O$_{6.95}$
has been studied by far-infared ellipsometry. A surprisingly lare number of
in-plane polarized modes are observed. Some of them correspond to pure
phonon modes. Others posses a large electronic contribution which strongly
increases in the superconducting state. The free carrier response and the
collective modes exhibit a pronounced a-b anisotropy. We discuss our results
in terms of a CDW state in the 1-d CuO chains and induced charge density
fluctuations within the 2-d CuO$_{2}$ planes.

\bigskip \noindent PACS Numbers: 78.30.-j, 74.25.Gz, 74.72.Bk, 74.25.Kc
\end{abstract}

\allowbreak

\newpage

The far-infrared (FIR) dielectric response of the cuprate high-T$_{c}$
superconductors (HTSC) has been extensively studied by conventional
reflection and absorption techniques \cite
{Schlesinger1,Renk1,Pham1,Bucher1,Basov1,Wang1,Homes1}. Nevertheless, some
important aspects of the FIR in-plane response remain unsettled, such as the
low-energy electronic excitations in the superconducting (SC) state or the
in-plane polarized IR-active phonon modes. The accuracy of the most popular
reflectance technique is limited by the reference problem (the absolute
value of the reflectivity, R, is required) and by the need of a
Kramers-Kronig analysis which involves extrapolation of R towards zero and
infinite frequency. Both problems become particularly important if R is
close to unity such as for the FIR in-plane response of HTSC. In this
manuscript we present experimental data of the in-plane dielectric response
of a detwinned YBa$_{2}$Cu$_{3}$O$_{6.95}$ single crystal which have been
obtained by spectral ellipsometry. This technique has the advantage that it
is self-normalizing and allows one to measure directly the complex
dielectric function, $\widetilde{\varepsilon }$=$\varepsilon _{1}$+i$\cdot
\varepsilon _{2}$ \cite{Azam1}. We find that the in-plane dielectric
response contains a surprisingly large number of collective modes, some of
which posses an unusually large spectral weight (SW) and exhibit pronounced
anomalies at the SC transition. These modes thus seem to involve collective
excitations of the charge carriers and of the SC condensate. The observed
in-plane anisotropy of the electronic background and the collective modes
cannot be consistently explained in terms of a superposition of independent
contributions from CuO$_{2}$ planes and 1-d metallic CuO chains. Our results
rather suggest that the 1-d CuO chains enhance (or possibly even induce) a
charge density modulation within the 2-d CuO$_{2}$ planes.

The ellipsometric measurements were performed at the U4IR and U10A beamlines
of the National Synchrotron Light Source (NSLS) in Brookhaven, USA. We used
a homebuilt setup attached to a conventional Fourier spectrometer \cite
{Henn1,Golnik1}. The high brilliance of the synchrotron allows us to study
small crystals in grazing incidence geometry ($\varphi $=85$%
%TCIMACRO{\UNICODE[m]{0xb0}}%
%BeginExpansion
{{}^\circ}%
%EndExpansion
\pm 1%
%TCIMACRO{\UNICODE[m]{0xb0}}%
%BeginExpansion
{{}^\circ}%
%EndExpansion
$) which is required for metallic samples with a Brewster-angle close to 90$%
%TCIMACRO{\UNICODE[m]{0xb0}}%
%BeginExpansion
{{}^\circ}%
%EndExpansion
$ \cite{Golnik1}. Since the dielectric response of HTSC is strongly
anisotropic we performed measurements for different configurations of the
principal axis (a, b, and c) with respect to the plane of incidence of the
light. The tensor components of $\widetilde{\varepsilon }$ have then been
obtained by standard numerical procedures \cite{Azam1}. We have previously
shown by measurements on gold films that a meaningful determination of $%
\widetilde{\varepsilon }$ for millimeter size metallic samples requires to
take diffraction effects into account \cite{Golnik1}. One of us (J.H.) has
studied these effects theoretically for a sample with infinite conductivity
and has developed an inversion procedure \cite{Humlicek1} which we have
successfully applied to data on small gold films \cite{Golnik1}. The same
procedure was used for the present data.

A Y-123 single crystal of size 2.8$\times $4$\times $0.6 mm$^{3}$ (a$\times $%
b$\times $c) was flux-grown in an Y-stabilized ZrO$_{2}$ crucible. It was
annealed for one week in flowing oxygen at 420 $%
%TCIMACRO{\UNICODE[m]{0xb0}}%
%BeginExpansion
{{}^\circ}%
%EndExpansion
$C resulting in T$_{c}$=91.5($\pm $0.8) K as determined by dc-SQUID
magnetometry. One ab- and one ac-face were polished to optical grade before
the crystal was mechanically detwinned as confirmed by inspection under a
polarizing optical microscope. Micro-Raman measurements confirm the
detwinning and the high quality of the surface \cite{Bernhardu}. They also
reveal a strong SC induced anomaly of the B$_{1g}$ mode at 340 cm$^{-1}$
which is characteristic for slightly overdoped Y-123 \cite{Altendorf1}.

Figures 1(i) and 1(ii) display the real part of the a-axis optical
conductivity, $\sigma _{1,a}$($\nu $), and of the dielectric function, $%
\epsilon _{1,a}$($\nu $) at 100 and 10 K (thin and thick solid lines). The
broad electronic features are similar to those reported for reflectivity
data \cite{Schlesinger1,Renk1,Basov1,Wang1,Homes1}. The NS the conductivity
consists of a Drude-like peak at low frequencies that merges with a broad
MIR-band which extends beyond the displayed range. In the SC state a
pronounced dip appears in $\sigma _{1,a}$($\nu $) with an onset around 950 cm%
$^{-1}$ and a sharp minimum around 320 cm$^{-1}$. For $\nu $%
%TCIMACRO{\TEXTsymbol{>}}%
%BeginExpansion
\mbox{$>$}%
%EndExpansion
1000 cm$^{-1}$ (inset of Fig 1(i)), $\sigma _{1,a}$ does not change between
100 and 10 K. This signals that the redistribution of electronic spectral
weight (SW) associated with the formation of the SC condensate involves an
energy scale of around 1000 cm$^{-1}$. Notably, a sizeable Drude-like peak
persists even in the SC state at 10 K. Its signature was observed also in
the reflection and absorption data \cite{Renk1,Pham1,Basov1,Wang1,Homes1},
by microwave measurements \cite{Hosseini1} and by mm-wave spectroscopy \cite
{Pimenov1}. Evidently, a significant fraction of quasiparticles (QP) does
not participate in the macroscopically coherent and thus loss-free response
of the SC condensate which is represented in $\sigma _{1}$ by a delta peak
at zero frequency ($\sigma _{1}\sim \nu _{p,SC}^{2}$*$\delta $($\nu $)), and
in $\varepsilon _{1}$ by a term ($\varepsilon _{1}\sim -\nu _{p,SC}^{2}$/$%
\nu ^{2})$, where $\nu _{p,SC}^{2}$ is the plasma frequency of the
condensate. Two kinds of processes can be expected to contribute to a finite
low-frequency conductivity. Firstly, scattering of charge carriers on
structural or magnetic defects leads to pair-breaking and thus gives rise to
unpaired QP's. This process should be very efficient for the cuprate HTSC
with their d-wave SC order parameter \cite{Lee1}. Secondly, dephasing
processes of the SC condensate can also give rise to a finite conductivity
at non-zero frequency. They can arise for example due to a spatial variation
of n$_{s}$ such as has recently been proposed based on THz-spectroscopy
measurements on Bi-2212 \cite{Corson1}.

In our ellipsometric spectra we also observe a number of well defined modes.
Our result for the SW, S=($\pi ^{2}\cdot c\cdot \varepsilon _{o}\cdot \nu
_{o}^{2})^{-1}\int \sigma (\nu )d\nu $, the resonance frequency (RF), $\nu
_{o}$, and the half-width at half-maximum, $\Gamma $, as obtained from fits
with a Lorentzian function, is listed in table I. The modes at 280, 360, and
600 cm$^{-1}$ have been previously identified \cite{Renk1,Homes1} and have
been assigned to phonon modes. The observation that these modes have a
similar SW as in the insulating parent compound YBa$_{2}$Cu$_{3}$O$_{6}$ 
\cite{Bauer1,Tajima1} has been interpreted in terms of a surprisingly poor
screening by the charge carriers of the CuO$_{2}$ planes which may be
related to an inhomogeneous charge distribution \cite{Homes1}. In addition
to these phonons we observe two more modes at 230 and 190 cm$^{-1}$ which
exhibit a remarkable behavior. Their RF's are close to the ones of some
in-plane polarized phonon modes in YBa$_{2}$Cu$_{3}$O$_{6}$ (see table I).
For both modes, however, the SW becomes extraordinarily large in the SC
state; at 10 K it exceeds the one of the pure phonon mode by more than an
order of magnitude. The SW of a phonon mode is, S=(4$\pi /V_{c}$)$\cdot
\lbrack \sum_{k}e_{k}^{\ast }\xi _{k}]^{2}/[\sum_{k}m_{k}\cdot \xi
_{k}^{2}\cdot v_{o}^{2}],$ where V$_{c}$ is the unit cell volume, e$%
_{k}^{\ast }$ and m$_{k}$ are the dynamic charge and the mass of the ion k
and $\xi _{k}$ its relative direct displacement. If the two present modes
were to be interpreted as pure phonons, their huge SW would imply that they
involve dynamic charges far in excess of the ones of the bare ions. They can
thus be rather ascribed to collective electronic excitations. The drastic SW
increase below T$_{c}$ furthermore suggests that they correspond to
excitations of the SC condensate. Above T$_{c}$ they are hardly detectable
and possibly correspond to the pure phonon modes. The eigenvectors of the IR
phonon modes in Y-123 have been previously obtained from lattice dynamical
calculations \cite{Kress1}. One a-polarized mode near 200 cm$^{-1}$ involves
the in-plane motion of O(2) and Cu(2) against O(3) and thus will couple
mainly to charge density oscillations within the CuO$_{2}$ planes. The other
relevant mode near 170 cm$^{-1}$ involves predominantly the motion of O(1)
against Cu(1) and therefore is sensitive to charge density oscillations
within the CuO chains.

We point out that our ellipsometric data provide sound evidence for the
low-energy modes. The spectra show the true dielectric function which is
corrected for anisotropy. The diffraction corrections are smooth functions
of d/$\lambda $ where d is the sample dimension in the plane of incidence
and $\lambda $ is the wavelength \cite{Golnik1} and thus cannot result in
any sharp features. Also, ellipsometry measures the real- and the imaginary
parts of $\widetilde{\varepsilon }$ independently. This allows us to check
the KK-consistency of the spectral features and thus to distinguish them
from structures related to noise or imperfections of the setup. It is also
very unlikely that the KK-consistent modes are caused by a damaged surface
layer. This conclusion is supported by our micro-Raman data which are even
more sensitive to the surface layer quality \cite{Bernhardu}. Irrespective
of their large SW, the signature of these modes is rather weak in the normal
incidence reflectivity, R, such as displayed in the inset of Fig. 1(ii).
Nevertheless, they seem to appear in some published spectra \cite
{Schlesinger1,Renk1,Basov1}. In particular, they have been identified in an
earlier absorptivity measurement \cite{Pham1}.

Figures 2(i) and 2(ii) display the spectra of $\sigma _{1,b}$($\omega $) and 
$\varepsilon _{1,b}$($\omega $) at 100 and 10 K (thin and thick solid
lines). The broad electronic response again agrees fairly well with the
previously reported one \cite{Renk1,Basov1,Wang1}. In particular, a sizeable
anisotropy is evident between the a- and the b-axis response with $\sigma
_{1,b}(\nu )$/$\sigma _{1,a}$($\nu $) $\sim 2$ in the NS, and n$_{s}^{b}$/n$%
_{s}^{a}$=($\lambda _{a}$/$\lambda _{b})^{2}\approx $(1700 \AA /1200 \AA )$%
^{2}\sim 2$ in the SC state, with the SC magnetic penetration depth, $%
\lambda _{a,b}$, derived from $\lambda _{a,b}^{-1}$=lim$_{\nu \rightarrow
0}\{\sqrt{4\cdot \pi ^{2}\cdot \nu ^{2}\cdot (1-\varepsilon _{1a,b}(\nu ))}%
\} $ (spectra are not shown). The enhanced free carrier response along the
b-axis direction was previously explained in terms of the fully oxygenated
1-d CuO chains which are thought to be metallic \cite{Schlesinger1,Renk1}
and to become SC due to proximity effect \cite{Basov1}. Our spectra also
reveal a large anisotropy of the energy scale of the SC condensation
process, i.e., of the range over which $\sigma _{1}$($\nu $) is reduced
below its value at T$_{c}$. In the b-axis spectrum this energy scale is
significantly larger (it exceeds the measured range) than in the a-axis one.
A similar trend appears in some reflectivity data \cite{Basov1,Wang1}. Such
a large anisotropy of the energy scale of the SC condensation process is
hardly consistent with the scenario of a proximity induced SC state in the
CuO chains which should lead to a decrease of the energy scale rather than
to the apparent increase.

Our b-axis spectra contain a surprisingly large number of collective modes.
At least nine modes are resolved in the 10 K spectrum at 140, 190, 230, 260,
295 340, 360, 480 and 560 cm$^{-1}$. Furthermore, the modes at 230, 290 and
480 cm$^{-1}$ exhibit a doublet structure. Table II lists the obtained
values for S, $\nu _{o}$, and $\Gamma $. Only the modes at 360 and 560 cm$%
^{-1}$ have a SW consistent with the assignment to pure phonon modes. The
large SW of the modes at 140, 190, 230, 260, 295, 340 and 480 cm$^{-1}$
suggests again that they involve electronic degrees of freedom.
Interestingly, these modes preserve a sizeable SW even in the NS, unlike the
two a-axis ones which almost disappear above T$_{c}$. For orthorhombic YBa$%
_{2}$Cu$_{3}$O$_{7}$ group theory predicts seven IR-active phonon modes for
both a- and b-polarization \cite{Kress1}. The multitude of the observed
b-polarized modes thus is indicative of an enlarged unit cell or a reduced
symmetry along the b-direction. It is well known that 1-d metals are
susceptible to structural and electronic instabilities, such as a transition
to a dimerized state or a charge density wave state (CDW) \cite{Dressel1}.
Indeed, recent NQR- and STM measurements indicate that the CuO chains of YBa$%
_{2}$Cu$_{3}$O$_{7}$ are in a CDW state \cite{Grevin1,Edwards1}. They
furthermore suggest that the CuO chains interact with the CuO$_{2}$ planes
and enhance (or possibly even induce) a charge density modulation therein in
the SC state \cite{Grevin1}. Such a scenario is consistent with our
observation of a-axis electronic modes which develop in the SC\ state and a
large number of b-axis polarized ones which occur already in the NS. The
scenario of a CDW state within the CuO chains, however, contradicts the
explanation of the in-plane anisotropy of the electronic conductivity in
terms of metallic CuO chains. A CDW\ is easily pinned by impurities, the CuO
chains consequently should be only poorly conductive. A broad hump around
250 cm$^{-1}$ which is very pronounced in the 10 K spectrum may well be
associated with localized charge carriers of the CuO chains. A similar
feature has previously been observed along the chain direction in Y-123 \cite
{Renk1} and Y-124 \cite{Bucher1}. The apparent contradiction can be resolved
if one associates the a-b-plane anisotropy of the free carrier response with
the CuO$_{2}$ planes themselves. Given that charge density fluctuations are
intrinsic to the CuO$_{2}$ planes, the interaction with the CuO chains may
indeed induce a strong anisotropy of the electronic response. The most
prominent example is the so-called stripe phase model which assumes a
pattern of alternating hole rich and hole poor 1-d stripes \cite{Emery1}.
Within such a scenario it is feasible that the interaction with the CuO
chains causes the stripes to become aligned along the b-direction, some
experimental evidence is reported in Ref. \cite{Mook1}. Such an arrangement
could indeed account for the observed anisotropy of the conductivity.
However, it would lead to a reduced symmetry along the a-axis direction
rather than along the b-axis one such as suggested by our experiments.
Therefore, it seems that one has to consider both effects, a CDW state
within the CuO chains and charge fluctuations within the CuO$_{2}$ planes
that develop due to the interaction with the 1-d CuO chains. In this context
it will be very interesting to learn whether similar collective modes and
anisotropy effects can be observed in other high-T$_{c}$ cuprates without
1-d CuO chains.

In summary, by ellipsometry we studied the in-plane FIR\ dielectric response
of detwinned YBa$_{2}$Cu$_{3}$O$_{6.95}$. We observe a surprisingly large
number of collective modes. Some of them correspond to in-plane polarized
phonons such as observed in undoped YBa$_{2}$Cu$_{3}$O$_{6}$. Others have an
unusually large spectral weight (SW) and therefore seem to involve
collective excitations of the charge carriers, possibly even of the
superconducting condensate. The collective modes and the free carrier
reponse exhibit a pronounced a-b anisotropy. We discussed our results in
terms of a CDW state in the 1-d CuO chains and induced charge density
fluctuations within the 2-d CuO$_{2}$ planes.

We thank D. B\"{o}hme and W. K\"{o}nig for expert technical help. T.H.
appreciates financial support by the AvH foundation.

\bigskip

Table 1: Values of the fitting parameters for the a-axis modes at 10 K
compared to the in-plane phonon modes of YBa$_{2}$Cu$_{3}$O$_{6}$ at 300 K
from \cite{Tajima1}$^{a}$ and \cite{Bauer1}$^{b}$.

\begin{tabular}{|c|c|c|c|c|c|}
\hline
\multicolumn{3}{|c|}{YBa$_{2}$Cu$_{3}$O$_{6.95}$} & \multicolumn{3}{|c|}{YBa$%
_{2}$Cu$_{3}$O$_{6}$} \\ \hline
S & $\nu _{o}({\rm cm}^{-1})$ & $\Gamma ({\rm cm}^{-1})$ & S$^{a/b}$ & ${\rm %
\nu _{o}^{a/b}(cm^{-1})}$ & $\Gamma ^{a/b}({\rm cm}^{-1})$ \\ \hline
0.7(0.04) & 595.6(0.2) & 25.1(0.6) & 0.6/0.4 & 588/595 & 26/28 \\ \hline
1.55(0.03) & 362.3(0.1) & 6.8(0.3) & 1.55/2.5 & 357/351 & 30/34 \\ \hline
2.3(0.2) & 279.4(0.4) & 41.2(2.2) & 0.85/1.4 & 231/246 & 14/8 \\ \hline
53.8(1.2) & 228.2(1.8) & 59.8(5.1) & 0.75/1.9 & 193/188 & 10/11 \\ \hline
18.5(0.5) & 189.5(0.1) & 10.2(0.8) & 1.05/2.1 & 118/116 & 9/8 \\ \hline
\end{tabular}

\bigskip

Table 2: Values of the fitting parameters for the b-axis modes.

\begin{tabular}{|c|c|c|c|c|c|}
\hline
\multicolumn{3}{|c|}{T=100K} & \multicolumn{3}{|c|}{T=10K} \\ \hline
S & $\nu _{o}({\rm cm}^{-1})$ & $\Gamma ({\rm cm}^{-1})$ & S & $\nu _{o}(%
{\rm cm}^{-1})$ & $\Gamma ({\rm cm}^{-1})$ \\ \hline
2.28(0.11) & 545.3(0.56) & 28.3(1.2) & 3.8(0.1) & 544.9(0.4) & 36(2) \\ 
\hline
11.5(0.4) & 480.8(0.28) & 53.4(2.5) & 13.6(0.3) & 476.8(0.2) & 46.6(1.3) \\ 
\hline
1.3(0.3) & 362.9(0.3) & 12.1(0.8) & 1.25(0.1) & 362.1(0.3) & 8.1(0.2) \\ 
\hline
2.5(0.2) & 343.3(0.2) & 6.8(0.5) & 4.4(0.2) & 343.8(0.2) & 8.2(0.2) \\ \hline
8.3(0.7) & 302.4(0.2) & 20(fix) & 14.7(0.6) & 299.6(0.6) & 19.8(0.8) \\ 
\hline
5.6(0.5) & 289.4(0.8) & 20(fix) & 15.1(0.9) & 287.0(0.7) & 21.3(1.0) \\ 
\hline
3.2(0.3) & 285.6(0.7) & 12.1(1.7) & 5.8(0.5) & 261.9(0.7) & 16.7(0.8) \\ 
\hline
10.5(0.9) & 238.8(0.9) & 20.2(2.1) & 48.9(2.6) & 236.1(1.0) & 29.6(2.9) \\ 
\hline
- & - & - & 41.4(2.1) & 220.1(0.7) & 24.1(3.2) \\ \hline
22.5(1.2) & 192.1(0.4) & 13.8(2.2) & 23.9(1.1) & 189.0(0.2) & 8.2(0.9) \\ 
\hline
15.4(1.4) & 174.6(0.8) & 13.4(1.7) & 124.0(5.2) & 139.3(2.3) & 43.9(8.5) \\ 
\hline
\end{tabular}

\section{Figure Captions}

\bigskip

Figure 1: a-axis component of the real part of (i) of the conductivity and
(ii) of the dielectric function of YBa$_{2}$Cu$_{3}$O$_{6.95}$ (T$_{c}$=91.5
K) at 100 K (thin solid line) and 10 K (thick solid line). Arrows mark the
collective modes. Inset: (i) $\sigma _{1a}$($\nu $) over an extended
spectral range and (ii) normal incidence reflectivity, R$_{a}$, as deduced
from the ellipsometric data.

Figure 2: b-axis component of the real part of (i) the far-infrared
conductivity and (ii) of the dielectric function of YBa$_{2}$Cu$_{3}$O$%
_{6.95}$ (T$_{c}$=91.5 K) at 100 K (thin solid line) and 10 K (thick solid
line). Dotted (solid) arrows mark the collective (doublet) modes. Inset: (i) 
$\sigma _{1b}$($\nu $) over an extended spectral range and (ii) normal
incidence reflectivity, R$_{b}$, as deduced from the ellipsometric data.

\bigskip

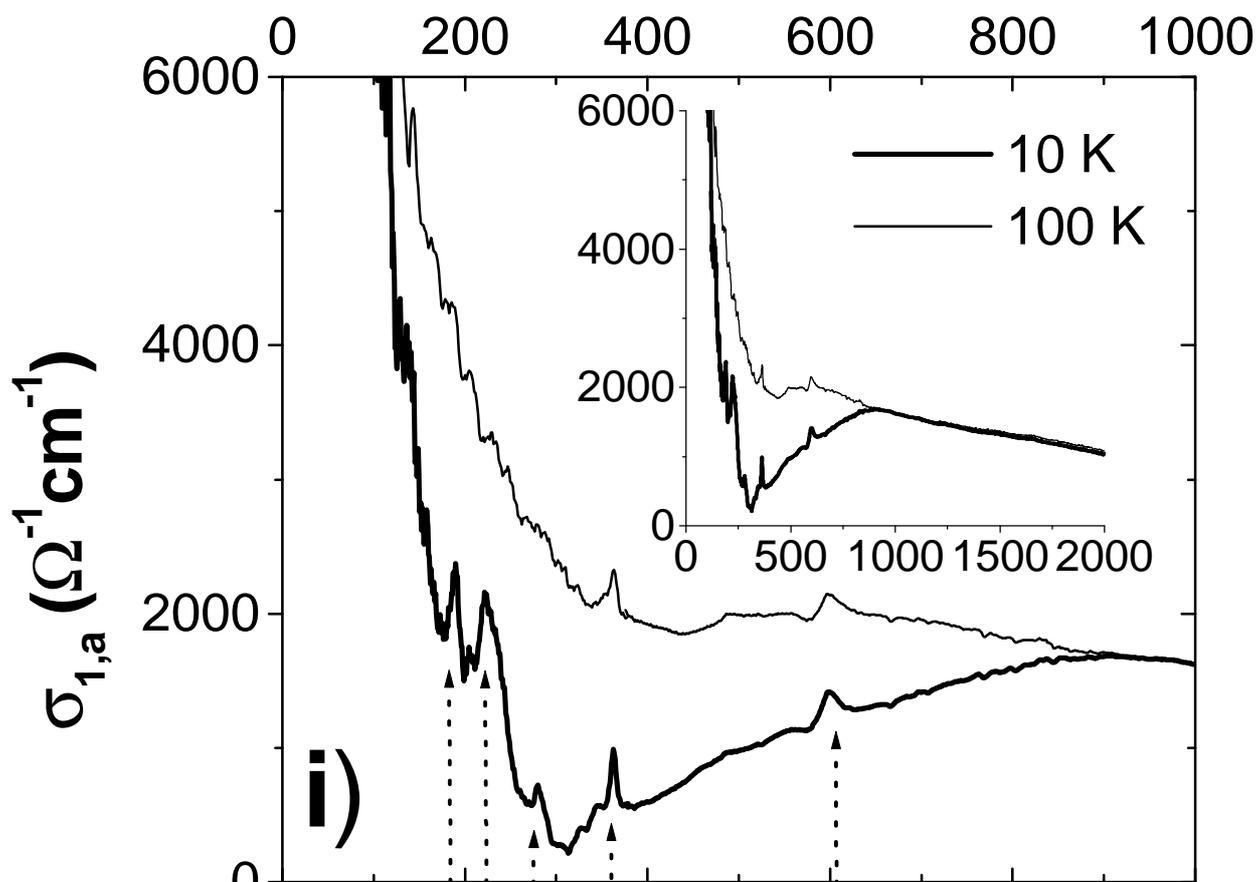
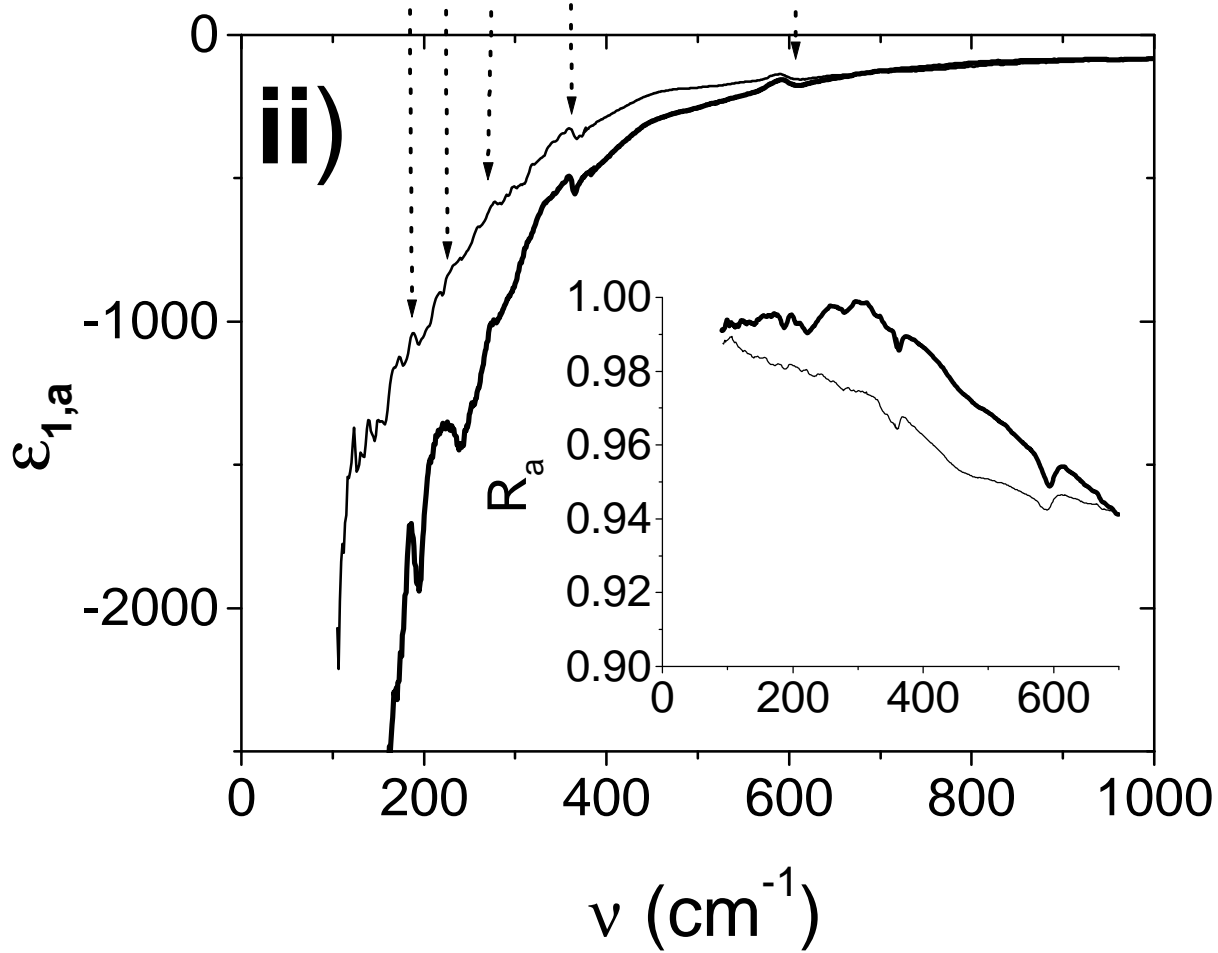
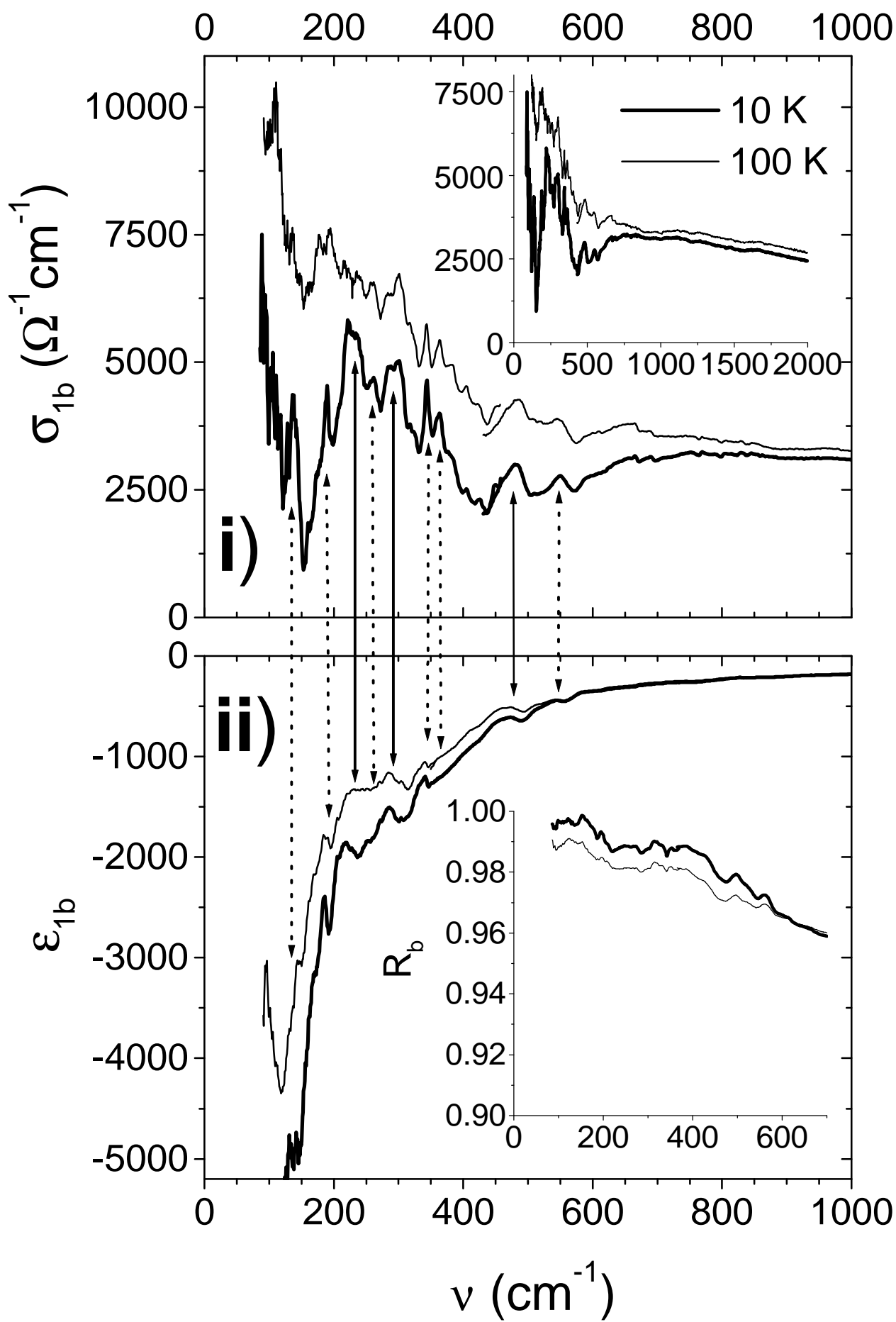
\end{document}